%% file: Paper.tex
\documentclass[acmlarge]{acmart}

\usepackage{booktabs} 

\usepackage[ruled]{algorithm2e} 

\SetAlFnt{\small}
\SetAlCapFnt{\small}
\SetAlCapNameFnt{\small}
\SetAlCapHSkip{0pt}
\IncMargin{-\parindent}

\setcopyright{usgovmixed}

\acmDOI{0000001.0000001}

\usepackage{makecell}

\usepackage{xspace}
\usepackage{graphicx}
\usepackage{epstopdf}
\usepackage{url}
\def\etal{\emph{et al.}}
\def\ie{\emph{i.e.}}
\def\eg{\emph{e.g.}}
\usepackage{csquotes}

\usepackage{textcomp}
\usepackage{gensymb}  

\usepackage{xspace}

\usepackage{multirow}
\usepackage{tabularx}
\usepackage{rotating}
\usepackage{amsmath}

\usepackage{graphicx}
\usepackage{subcaption}
\usepackage{mwe}

\usepackage{booktabs} 

\usepackage{color}
\definecolor{Orange}{rgb}{0.9,0.5,0}
\definecolor{NavyBlue}{rgb}{0.1, 0.4, 0.8}
\definecolor{Magenta}{rgb}{0.8, 0.1, 0.6}
\definecolor{Green}{rgb}{0.1, 0.8, 0.3}
\definecolor{DarkGreen}{rgb}{0.0, 0.7, 0.2}
\definecolor{Brown}{rgb}{0.4, 0.3, 0.1}
\definecolor{Burgundy}{rgb}{0.5, 0.0, 0.13}
\definecolor{BrightCerulean}{rgb}{0.11, 0.67, 0.84}
\definecolor{BlueViolet}{rgb}{0.33,0.1,0.5}

\usepackage{siunitx}

\begin{document}

\title{Finding Dory in the Crowd: Detecting Social Interactions using Multi-Modal Mobile Sensing}

\author{Kleomenis Katevas}
\affiliation{%
  \institution{Imperial College London}
  \country{UK}
}

\author{Katrin H\"ansel}
\affiliation{%
  \institution{Queen Mary University of London}
  \country{UK}
}

\author{Richard Clegg}
\affiliation{%
  \institution{Queen Mary University of London}
  \country{UK}
}

\author{Ilias Leontiadis}
\affiliation{%
  \institution{Telef\'onica Research}
  \city{Barcelona}
  \country{Spain}
}

\author{Hamed Haddadi}
\affiliation{%
  \institution{Imperial College London}
  \country{UK}
}

\author{Laurissa Tokarchuk}
\affiliation{%
  \institution{Queen Mary University of London}
  \country{UK}
}

\begin{abstract}
\input{00_Abstract.tex}
\end{abstract}

%
%


%
%


\maketitle


\renewcommand{\shortauthors}{Katevas~\etal}

\input{01_Introduction}
\input{02_Related}
\input{03_Proximity_Estimation.tex}
\input{04_Experimental_Setup.tex}
\input{05_Detecting_Interactions.tex}
\input{08_Results.tex}
\input{09_Discussion_and_Implications.tex}
\input{10_Conclusion}

\section{Acknowledgements}
The authors wish to thank the participants of this study. This work is supported by funding from the UK Defense Science and Technology Laboratory.

\bibliographystyle{abbrv}
\bibliography{references}

\end{document}

%% file: 00_Abstract.tex
Remembering our day-to-day social interactions is challenging even if you aren't a blue memory challenged fish. The ability to automatically detect and remember these types of interactions is not only beneficial for individuals interested in their behavior in crowded situations, but also of interest to those who analyze crowd behavior. Currently, detecting social interactions is often performed using a variety of methods including ethnographic studies, computer vision techniques and manual annotation-based data analysis. However, mobile phones offer easier means for data collection that is easy to analyze and can preserve the user's privacy. In this work, we present a system for detecting stationary social interactions inside crowds, leveraging multi-modal mobile sensing data such as Bluetooth Smart (BLE), accelerometer and gyroscope. To inform the development of such system, we conducted a study with 24 participants, where we asked them to socialize with each other for 45 minutes. We built a machine learning system based on gradient-boosted trees that predicts both 1:1 and group interactions with 77.8\% precision and 86.5\% recall, a 30.2\% performance increase compared to a proximity-based approach. By utilizing a community detection-based method, we further detected the various group formation that exist within the crowd. Using mobile phone sensors already carried by the majority of people in a crowd makes our approach particularly well suited to real-life analysis of crowd behavior and influence strategies.

%% file: 01_Introduction.tex
\section{Introduction}
\label{sec:intro}

The ability to automatically detect social interactions in unorchestrated scenarios is highly sought after in many areas including social and behavioral science, crowd management, and targeted advertising. This ability would facilitate a wide range of technologies, for example: (i) crowd reconfiguration in evacuation management, providing instructions strategically to groups is more efficient than to individuals and avoids different members of a group being sent conflicting instructions; (ii) networking analytics, allowing individuals to trace their interactions in networking events (instead of exchanging business cards) and providing analytics to event organizers to optimize and monetize events; (iii) targeting advertisements to groups. 

There have been many attempts for detecting social interactions automatically, primarily from video analysis. Most of the initial works use resource-hungry computer vision techniques~\cite{Hung:2011,Cristani:2011wp,Bazzani:2012}. Other approaches use custom-made wearable hardware that use sensors such as infrared light~\cite{Choudhury:2003,Olguin:2008,Huang:2014:Opo,Montanari:2018}, accelerometer~\cite{Huang:2014:Opo}, microphone~\cite{Choudhury:2003,Huang:2014:Opo} and Bluetooth~\cite{Huang:2014:Opo}. These works report reasonable accuracy but are expensive and problematic to scale in larger environments.

Smartphones and their wide range of embedded sensors enable researchers to explore social interactions in an automated way that depends entirely on the use of mobile sensing technology~\cite{Palaghias:2015,Katevas:2016:HASCA,Zhang:2016}, without the need for additional wearable equipment or computer vision systems. Mobile sensing-based solutions are also easier and more cost efficient to deploy in unknown or new spaces as they only rely on the users' own hardware. Early systems that use mobile sensing report accurate results, but focus on detecting one-to-one social interactions~\cite{Palaghias:2015}. Furthermore, they rely on pre-trained models that only work with specific mobile devices~\cite{Palaghias:2015}. Others are restricted to controlled only environments~\cite{Katevas:2016}, a situation that only covers a subset of the formations that occur in a natural setting, or use the phone's microphone for detecting body distances~\cite{Zhang:2016}, an approach that raises concerns about the user's privacy.

In this paper, we investigate an approach for detecting social interactions in a natural, non-artificial social setting. We focus on the interactions that usually happen in social gatherings or networking events (\eg, conferences, exhibition etc.) where people form standing interactions with two or more participants. We built a machine learning system based on gradient-boosted trees to detect both 1:1 and group interactions in a short granularity of 1 second window. We then use a community detection algorithm based on graph theory to detect the various group formation that exist within the crowd. We evaluate our system in a case study with 24 participants interacting together for 45 minutes. We tested two different approaches for inferring whether a group of people are close enough that a social interaction is feasible: (i) using high-performance, long-range beacons installed in the ceiling of the room, and (ii) without any fixed infrastructure. Notice that due to software limitations, the phones were not able to transmit beacons when the device is locked. Therefore, we ended up using coin-shaped beacons as a wearable device that simulates the smartphone's Bluetooth broadcasting function.

The main contributions of this work are:

\begin{itemize}

\item A machine-learning-based approach that predicts social interactions relying on data collected via the mobile phone, achieving a 77.8\% precision and 86.5\% recall, a 30.2\% performance increase in terms of Average Precision compared to a proximity-based approach.

\item A graph theoretical solution applied in the context of detecting social interactions that is capable of detecting the different types of group formations that exist within the crowd.

\item A proper evaluation in a natural (not artificially created) environment in three ways: (a) \emph{link-level}, when an interaction between a pair of participants exist, (b) \emph{node-level}, where a participant belongs to the correct interactive group, and (c) \emph{group-level}, where a group of people is detected to include all participants correctly.
\end{itemize}

We further contribute and share a freely available dataset with unconstrained natural one-to-one and group interactions in varying sizes. To our knowledge, we are the first to present a system that automatically detects interactive groups of various sizes using privacy-aware mobile sensor data and evaluated in a natural setting.

%% file: 02_Related.tex
\section{Related Work}
\label{sec:related}

Sensing social interactions has been an interest in researchers and stakeholders in various fields. 
One of the main objectives of architects, designers and organizational researchers is to promote, design and measure meaningful and productive interactions in office workspaces. Face-to-face interactions are hereby considered the most valuable form of communication~\cite{Pentland:2012-building-teams}.
Research has shown that meaningful interactions, like \eg, influenced by the office layout, can boast productivity, well-being or team cohesion~\cite{becker1995workplace,blok2009effects}.
While some work focuses on agent-based modelling of office interactions~\cite{Langevin2016-agentoffice}, traditional approaches include manual observations and annotations in real-world scenarios. These have the drawback of being time-consuming and can influence office worker behaviour while being monitored. 
The use of automated sensing approaches can support easier, larger scale evaluation of office interactions, dynamics and well-being~\cite{Mathur:2015-quant-workspace}. 
Hereby, sensing approaches within the office has focused on using customary hardware, such as wearable badges (\eg, Sociometer~\cite{Choudhury:2003}, SocioPatterns~\cite{Brown:2014-serendipity-workplace}, or Protractor~\cite{Montanari:2018}), mobile phone Bluetooth sensing (\eg, Efstratiou~\etal~\cite{Efstratiou2012-sensibility-mobile}) or hybrid approaches (\eg, Matic~\etal~\cite{Matic:2012}).
Office and workspaces follow special rules as spaces are disrupted by objects such as desks, furniture etc. These objects can sometimes act as facilitators of interactions, such as kitchen areas or photo copiers~\cite{Fayard:2007-office-copier}.
Further, office spaces are not densely populated, as for example event spaces like conferences or exhibitions.
On the contrary, a confined office space offers large control over the environment and can allow the placement of additional ambient sensors within the space~\cite{Mathur:2015-quant-workspace}.
Additionally, these semi-private spaces are usually frequented by a specific subset of co-workers which makes the distribution of custom-made sensing badges feasible.

Crowd and interaction sensing on open -- large spaces with a fluctuating number of people can benefit from using existing hardware such as the phones owned by the users. This reduces the need to deploy custom hardware and devices which have to be managed centralized and handed out to users. However, challenges arise in terms of different hardware (\eg, sensors) in mobile phones. Previous works trained device-specific models that use embedded sensors such as WiFi~\cite{Matic:2010,Matic:2012} or Bluetooth~\cite{Palaghias:2015} to detect the user proximity. Crowded spaces with people standing close require fine-grained sensing mechanisms or novel approaches, \eg, Zhang~\etal~\cite{Zhang:2016} used audio sensing to detect turn-taking in conversations as additional features.

Computer vision has also been applied into the area of detecting social interactions. Hung \& Krose~\cite{Hung:2011} proposed a solution to detect face-to-face social interactions from manual-annotated video footage. In their study they used information about the proximity between people, as well as the body orientation to identify interactions with an accuracy of 92\%. In a similar research, Cristani~\etal~\cite{Cristani:2011wp} suggested a system that also detects interactions using information about people's position and head orientation from visual cues. They also reported comparable results of 89\% accuracy. Both works assume that the information of related proximity between participants or absolute position in the space, as well as the body or head orientation is known, either using manual annotations or computer vision techniques. Bazzani~\etal~\cite{Bazzani:2012} presented a more sophisticated computer vision-based approach for tracking groups using a join individual-group tracking framework. They evaluated their approach in a video-based dataset recorded on an outdoor area where people usually meet during coffee breaks, achieving a group detection performance of 71.70\% accuracy.
These visual approaches on proximity and orientation of people can be transferred to on-body-sensing mechanisms. 

Using wearable and mobile sensors opposed to environmental and vision techniques offers more flexible and location independent identification of interactions. 
One of the first attempts to identify stationary, face-to-face interactions in an automated way is the \emph{Sociometer} by Choudhury and Pentland~\cite{Choudhury:2003}, a wearable device that can be placed on each person's shoulder and identify other people wearing the same device using Infrared (IR) sensors. In addition, it is equipped with an accelerometer sensor to capture motion as well as a microphone to capture speech information.
Olguin~\etal~\cite{Olguin:2008} developed a successor of the Sociometer, called the \emph{Sociometric} badge, that is smaller is size and includes Bluetooth, IR, microphone and accelerometer sensors. Huang~\etal~\cite{Huang:2014:Opo} designed a low-power wearable device capable of detecting human interactions using ultrasonic signal. They evaluated their device in a series of human experiments with both sitting and standing interactions. Montanari~\etal~\cite{Montanari:2018} created a wearable device named \emph{Protractor} that uses near-infrared light to monitor the user proximity and relative body-orientation. The device was evaluated in a group-collaborative task where 64 participants split into groups of four were asked to collaborate with each other to build a construction made of spaghetti and plastic tape.
Note that even though the evaluation of this work is focused on the social behavior of an existing group that is interacting, Protractor could also be used to detect  human interactions within crowds by using the estimated proximity and relative orientation between participants. The used sensors hereby are accurate, but require a special placement of the sensor and no occlusion from \eg, clothing.

Opposed to deploying custom made sensors and badges, novel work focused on leveraging off-the-shelf devices and smartphone sensors. Palaghias~\etal~\cite{Palaghias:2015} presented a real-time system for recognizing social interactions using smartphone devices. Using the RSSI of Bluetooth Classic radios and a 2-layer machine learning model, they classified the proximity between two devices into three interaction zones, based on the theory of Proxemics~\cite{proxemics}: \emph{public}, \emph{social} and \emph{personal}. In addition, they used an improved version of \emph{uDirect} research~\cite{uDirect} that utilizes a combination of accelerometer and magnetometer sensors to estimate the user's facing direction with respect to the earth's coordinates. This work reported results of 81.40\% accuracy for detecting social interactions, with no previous knowledge of the device's orientation inside the user's pocket. However, this work has been evaluated in a limited dataset with eight participants while an observer was keeping notes that were later used as ground truth. Moreover, it is only capable of detecting one-to-one social interactions using a specific device model (HTC One S) and has not been evaluated in scenarios of interactions with dynamic sizes. Finally, it assumes that a Bluetooth connection is maintained between devices for continuously monitoring the RSSI, having an impact on the device's battery. Zhang~\etal~\cite{Zhang:2016} developed a system that detects social interactions in the context of encountering with the use of audio sensing. They first used a combination of the smartphone's accelerometer, microphone and speaker, and with the use of inaudible acoustic signals they detected when two people approach and stop in front of each other. Next, they applied voice profiling on the audio recordings to confirm if the pair is engaged into an actual conversation. They evaluated their approach in a real-world use case with 11 participants for 1 hour using self-reported questionnaires at the end of the study as ground truth. The evaluation of this work that reports 6.9\% false positives and 9.7\% false negatives, was conducted over the complete case study (\ie, who met with whom during the event) instead of a more fine-grained evaluation over short windows (\eg, per second). Thus, it is not capable of capturing information such as the duration of an interaction, or more advanced crowd dynamics such as type of group formations that were conducted over time. Moreover, such approach requires a continuous audio recording from each user's smartphone, a process that raises ethical and privacy concerns when using it in real-world scenarios. Katevas~\etal~\cite{Katevas:2016:HASCA} presented a simplistic proximity-based approach for detecting stationary interactions in planned events, using the interpersonal proximity estimated by the device's Bluetooth Smart sensor. They evaluated the social interactions that took place in a controlled environment with six participants for four minutes, reporting a performance of 90.9\% precision and 92.4\% recall. This work was evaluated in a limited dataset (approx. 5 minutes long) with artificially created interactions instructed by the designer of the study. Moreover, the proximity-based algorithm they used is similar to the baseline used in this work.

In summary, using custom made hardware and badges may provide more control over the sensors, allow fine-grained and purpose-focused sensing. While this approach is suited for scenarios with limited users due to the cost and effort to build and distribute a wearable device, modern smartphones offer a flexible solution. Previous works have focused on detecting one-to-one interactions only, evaluated in controlled environments or depend on privacy-sensitive data such as voice recordings. In this work, we suggest an approach that uses a smartphone device and a series of privacy-aware sensors (\ie, Bluetooth Smart, accelerometer and gyroscope) to detect interactions of varying sizes that usually happen in social gatherings or networking events, evaluated in a natural, non-artificial social setting.

%% file: 03_Proximity_Estimation.tex
\section{Interpersonal Proximity Estimation}
\label{sec:proximity_estimation}

In this section we present two experiments that evaluate the use of BLE-based beacon technology to estimate whether two participants are close enough for a social interaction to be feasible. The aim is to investigate whether the beacon's RSSI on a custom \emph{Broadcasting Power} configuration can be a good predictor for a supervised machine learning classifier.

There have been several ways of estimating the distance between devices using wireless sensors such as \emph{Time of Arrival}, \emph{Time Difference of Arrival}, \emph{Angle of Arrival} and using the \emph{RSSI}. Currently, the only method that is applicable in smartphones is the RSSI of either the Bluetooth or the WiFi sensor. In the past, researchers have used the RSSI of Bluetooth~\cite{Liu:2014,Hu:2013,Palaghias:2015}, WiFi~\cite{Matic:2010} or even a combination of them~\cite{Banerjee:2010} by measuring the RSSI of every wireless sensor available in range and comparing it with a \emph{Measured Power} constant that indicates the signal strength (in dBm) at a known distance (usually $1m$). In 2010, the Bluetooth Special Interest Group released Bluetooth v4.0 with a Low Energy feature (BLE) that was branded as Bluetooth Smart. Bluetooth Smart is low cost for consumers, has low latency in communications ($6 ms$) and is power efficient. Moreover, it supports a low energy advertising mode where the device periodically broadcasts specially formatted advertising packets to all devices in range with a customizable sample rate of approximately $3 Hz$. This packet can include 31 bytes of information, such as a unique ID for each user, but also the measured power constant that was mentioned above. The advantage of using this technology for proximity estimation is that each manufacturer can configure the device to use its own pre-calibrated measured power constant, making the proximity estimation more accurate and device-type independent. In addition, devices do not need to maintain a connection with each other in order to measure the RSSI, having a minimum impact on the device's battery life~\cite{Katevas:2016}. In its latest version, marketed as Bluetooth~5~\cite{bluetooth}, the sensor provides additional benefits including longer range (x4) and longer capacity in the advertising packet (x8).

Apple developed a proprietary protocol based on Bluetooth Smart, branded as \emph{iBeacon} and supported it as of iOS~7 (June 2013) in all mobile devices with Bluetooth~4.0 or greater (iPhone~4 or newer)~\cite{apple_ibeacon}. The specification of iBeacon advertising packet includes: one 16-byte \emph{Universally Unique Identifier (UUID)} used to separate beacon applications, two 2-byte unsigned integer identifiers named \emph{Major}, which separates beacon groups (e.g. on the same venue or floor), and \emph{Minor}, which separates individual beacons within the group, and one 1-byte \emph{Measured Power} value used as an RSSI reference at $1m$ distance. The remaining available bytes are used as a static prefix and cannot be customized by the developer. An iOS application can register to monitor for beacons of specific UUIDs and estimate its proximity whenever a beacon exists within range. The app can also advertise iBeacon packets, however, only while the device is unlocked (\ie, in-use while the screen is on) and the app remains in the foreground, a restriction applied by the mobile operating system. Furthermore, it is not possible to customize the \emph{Broadcasting Power} of the Bluetooth sensor, using the maximum power by default. Note that even though iBeacon is an Apple product for iOS devices, it is possible to scan or broadcast as an iBeacon from Android platform using third-party libraries\footnote{https://github.com/AltBeacon/android-beacon-library}. Similar restrictions have been added in Android platform with the release of Android v8.0 (Oreo), restricting apps to execute long-running services in the background\footnote{https://developer.android.com/about/versions/oreo/background}.

To overcome these limitations, we use wearable beacons (\ie, RadBeacon Dot from Radius Networks\footnote{https://www.radiusnetworks.com}) to broadcast a beacon signal while a sensor data collection app is running as a background process on each user's smartphone. This also allows the customization of the broadcasting power of each beacon, achieving better accuracy in estimating the social space of each participant as shown below. What we propose is compatible with the current hardware available in the majority of iOS and Android devices (\ie, with Bluetooth v4.0 or greater), and the restriction is software-based from the mobile operating system.

While previous evaluations report RSSI results from setups with beacons and phones mounted on tripods~\cite{Katevas:2016:HASCA}, or water bottles~\cite{Palaghias:2015} to simulate the effect of body water on the beacon's RSSI, neither captured the effect of human posture or blockage by body parts. Moreover, the signal is not only affected by the body's water, but also the electric properties of human tissues (muscle, fat and skin)~\cite{rapinski2016influence}. In this experiment, actual participants were used as a more realistic environmental setting.

\subsection{Effect of Broadcasting Power in the RSSI}

In order to evaluate the optimal broadcasting power setting for detecting if a pair is within a social enabled zone, we conducted a short experiment. Two participants were recruited: P1, male with height $1.79m$ and weight $73kg$, and P2, male with height $1.83m$ and weight $87kg$. P1 served as the broadcaster and was equipped with eight coin-shaped beacons of the same type. Each beacon was configured to a different broadcasting power setting. P2 had the role of the receiver and had an iPhone SE device placed in one pocket. A mobile sensing app was used to collect iBeacon Proximity data from all eight beacons, for 30 seconds at 15 distances from $0.25$ to $4.00$, every $0.25m$.

\begin{figure}[t!]
\centering
\includegraphics[width=\linewidth]{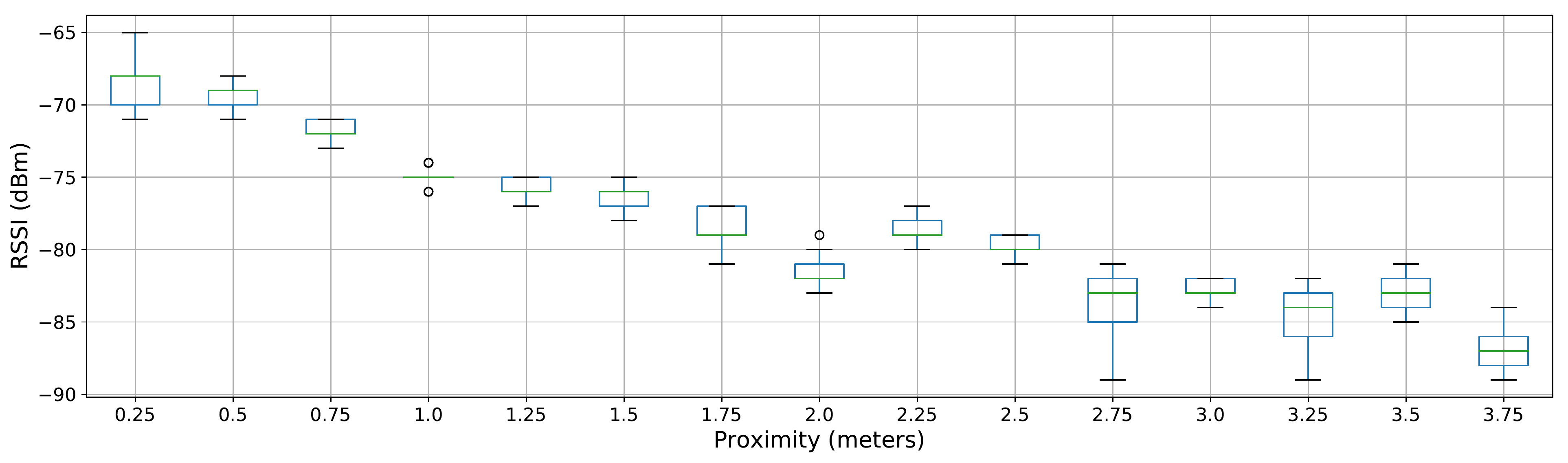}
\caption{iBeacon RSSI at varying proximity at minimum broadcasting power (~$-18 dBm$).}
\label{fig:beacon_proximity_figure_rssi}
\end{figure}

Our results show that each beacon, due to its configuration, has a different RSSI range and a unique pattern. For example, for the highest $+3 dBm$ power, it was challenging to differentiate between distances $0.75$ and $1.25$, or $1.00$ and $1.50$. Most signals greatly fluctuate, especially the longer distances (>$1.75m$). We chose the minimum broadcast power (\ie, $-18 dBm$) as it clearly separates the RSSI in distances until $1.5m$ (see Figure~\ref{fig:beacon_proximity_figure_rssi}). Moreover, the signal looks relatively smooth compared to the others, which should aid classification in the distances of interest. Similar choice was made in~\cite{Matic:2012} where the device's WiFi sensor in lowest power was used for the detecting social interactions.

\subsection{Effect of Body Orientation in the RSSI}

The low frequency of Bluetooth results in RSSI measurements that are highly affected by the human body. A second experiment was conducted to report how the RSSI is affected by relative body orientation and whether there are distinctive patterns that a machine learning classifier could benefit from. P1 and P2 were asked to stand facing each other at $1m$ distance and engage in a conversation. P1 was the broadcaster having two coin-shaped beacons configured to $-18dBm$ broadcasting power, one placed in each of his pockets. P2 was the receiver having two iPhone~SE phones, one in each of his pockets. An app collected data for $30 sec$ on all 64 combinations of different orientations, with a resolution of 45\textdegree. For each $30 sec$ window, data was collected for all four combinations of device placement (left/right pocket).

\begin{figure}
    \centering
    \begin{subfigure}{110pt}
        \includegraphics[width=\linewidth]{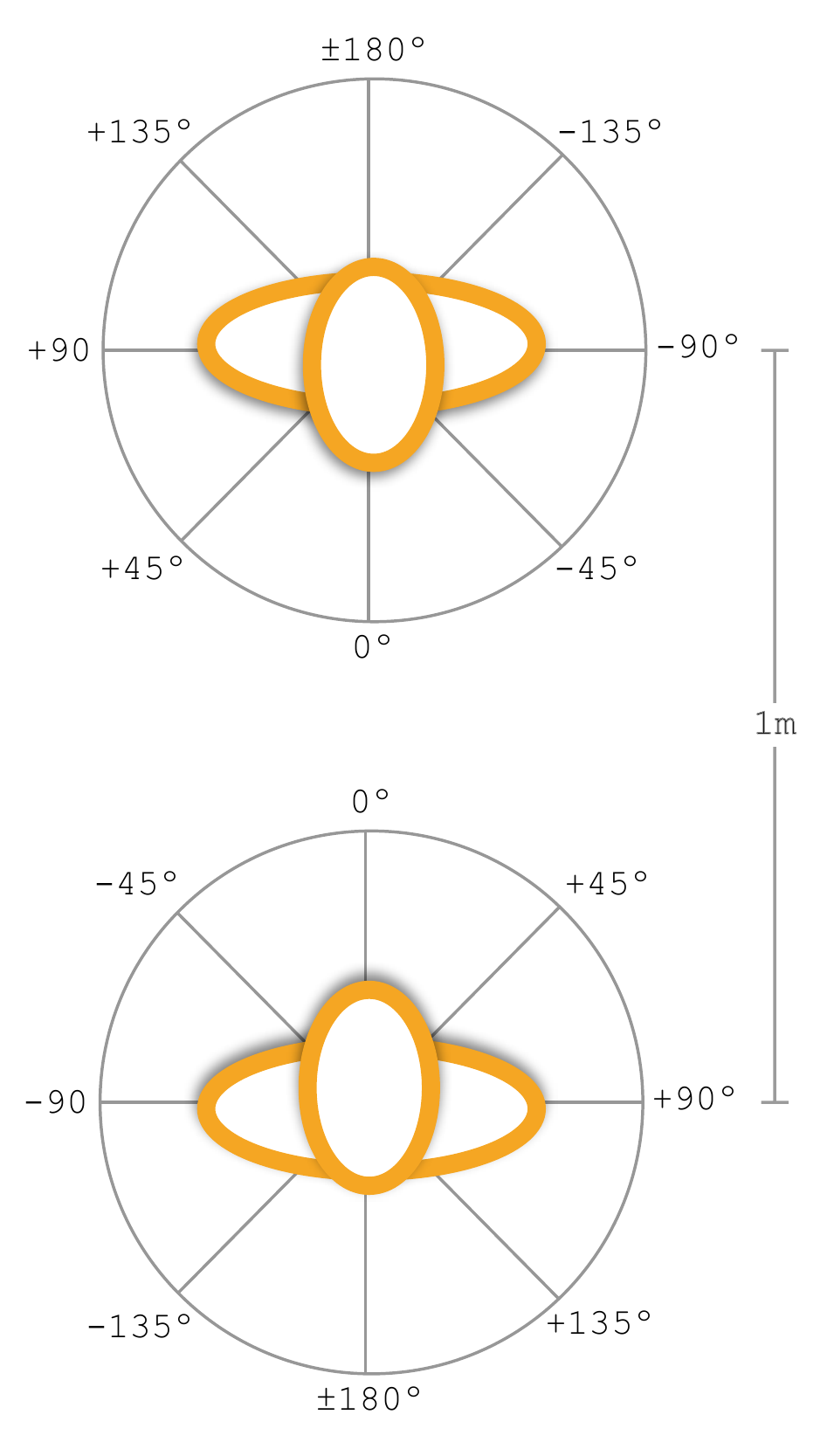}
        \caption{}\label{fig:beacon_orientation_configuration}
    \end{subfigure}
    \qquad
    \begin{subfigure}{200pt}
        \includegraphics[width=\linewidth]{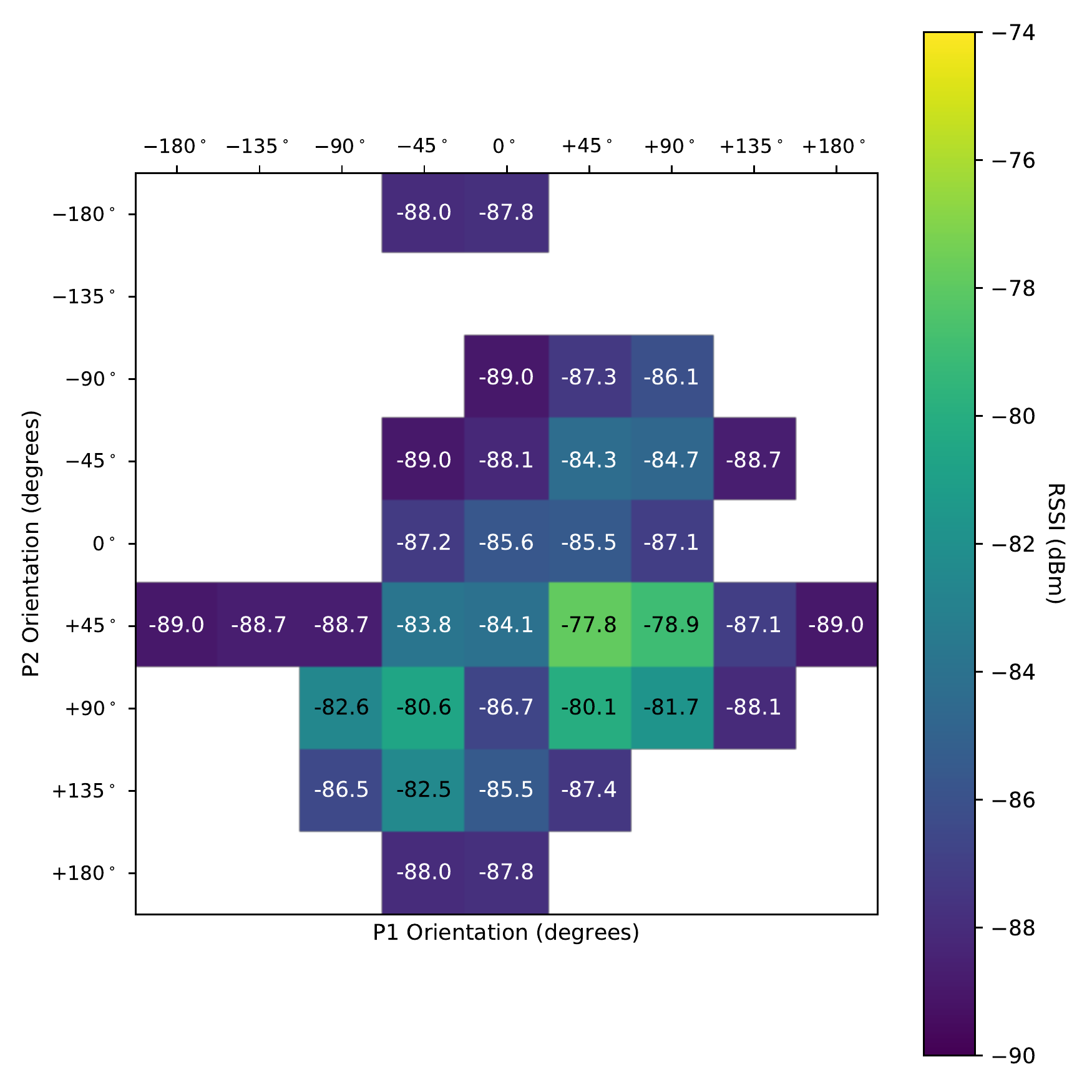}
        \caption{}\label{fig:beacon_orientation_heatmaps}
    \end{subfigure}
    \caption{Heat map visualization of the mean RSSI of over different relative orientations of the two participants, having the devices placed at their left pocket. Blank entries indicate no data due to the low \emph{Broadcasting Power} configuration used.}
    \label{fig:beacon_orientation}
\end{figure}

Figure~\ref{fig:beacon_orientation} shows that the RSSI varies based on the relative orientation of the two participants. There are orientations that the device could not receive any signal from the beacon as the body effectively blocked the broadcaster's signal. This is a desirable result as social interactions are not possible at such orientations. Other important discoveries are (a) the RSSI differs based on the configuration of in which pocket each device was placed (left or right), and (b) the measurements per configuration are not symmetric. For instance, someone would expect that in a scenario where P1 is 0\textdegree and P2 90\textdegree, the measured RSSI would be similar to the symmetric scenario where P1 is 90\textdegree and P2 90\textdegree. The asymmetry in the results can be attributed to the configuration, using P1 as the beacon broadcaster and P2 as the beacon receiver. This suggests that knowing this configuration (\ie, in which pocket the user placed his/her phone) would result in better accuracy.

%% file: 04_Experimental_Setup.tex
\section{Experimental Setup}
\label{sec:methodology}

In order to identify and evaluate the sensors needed for detecting stationary interactions in a natural setting, data was collected from participants during a social networking event. This section includes a description of the participants (Section~\ref{sec:case_study:Participants}), the procedure followed (Section~\ref{sec:case_study:Procedure}), as well as the sensor data collected (Section~\ref{sec:case_study:Sensors}).

\subsection{Participants}
\label{sec:case_study:Participants}

37 potential participants were recruited via email and flyers; 24 of those took part in the actual study of which 9 were male and 15 female, with average weight $63.75 kg$ ($\pm18.02$), and average height $167.21 cm$ ($\pm9.11$). Participants were selected based on mobile phone model (iPhone~4 or higher) and operating system version  (iOS~7 or higher) and availability of the iBeacon sensor. Two devices experienced errors during the study (\ie, Bluetooth Smart sensor reported an internal error and did not collect data) and were excluded from the data analysis, resulting into 22 valid participants.

\subsection{Procedure}
\label{sec:case_study:Procedure}

\begin{figure}[t!]
    \centering
    \includegraphics[width=300pt,trim={1.5cm 1.5cm 1.5cm 1.5cm}]{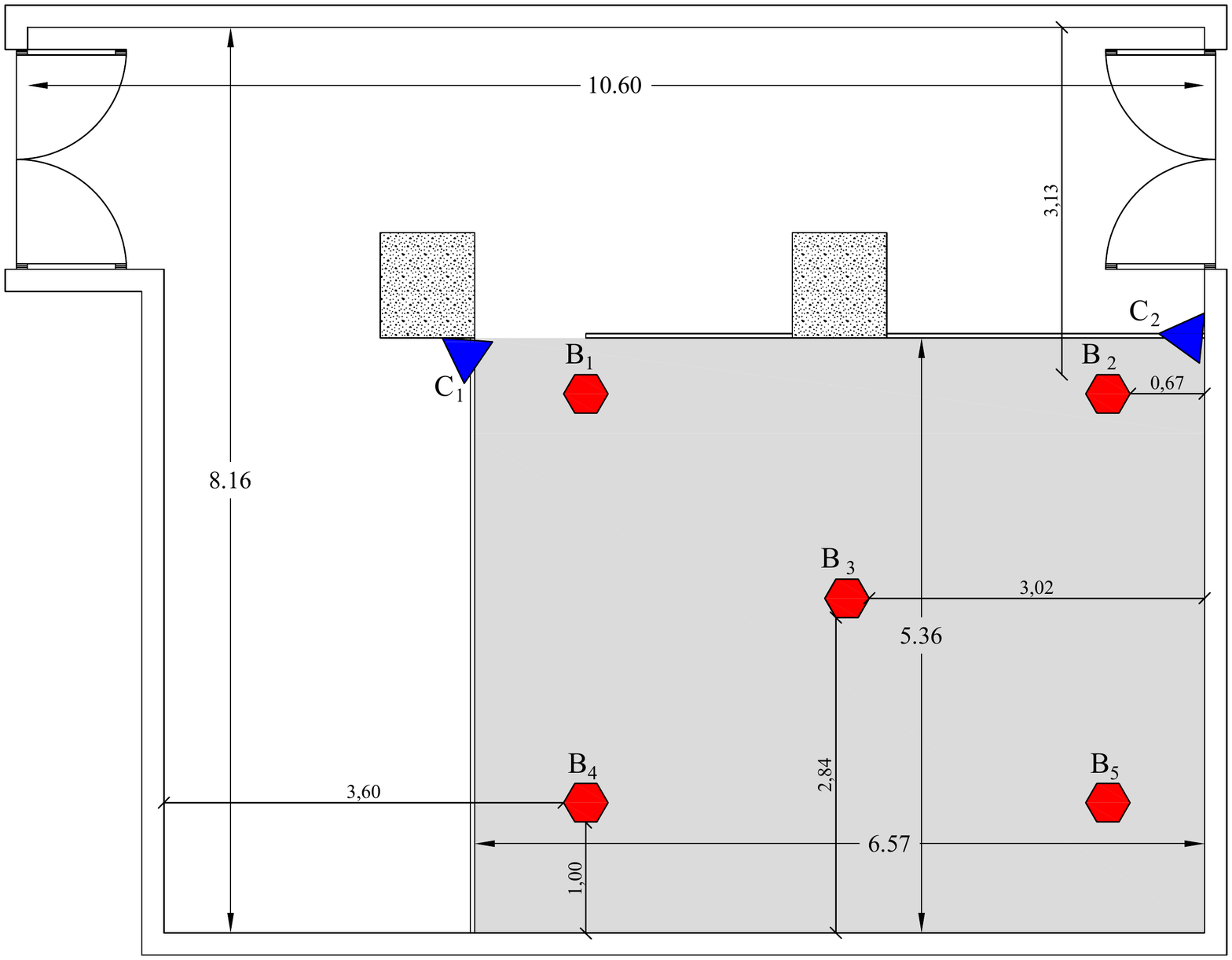}
    \caption{Floor plan of experimental location. The space contains two cameras ($C_i$) in blue, five ceiling beacons ($B_i$) in red and the interaction space is highlighted in grey.}
    \label{fig:floor_plan}
\end{figure}

Participants installed a sensor data collection app, based on SensingKit for iOS v0.5 continuous sensing framework~\cite{Katevas:2016}. The app automated the sensor calibration, participant registration and data collection. Participants were invited to an indoor location with the floor plan given in  Figure~\ref{fig:floor_plan}. The space is $10.60\times8.16$ meters, with $3.90 m$ height; it is suitable for such type of experiments not only due to its isolation from outside noise and environmental factors, but also due to it being a natural space often used for social events and performances. It provides a DMX lighting rig installed in $3.27 m$ height which was used to fix two HD cameras ($C_i$) recording video (but not audio) in $25fps$, covering an area of $6.57\times5.36$ meters (highlighted in grey in Figure~\ref{fig:floor_plan}). These videos were annotated to provide the ground truth for social interaction (see Section~\ref{sec:data_analysis:Ground_Truth}). This area was restricted using plastic dividers of $1.94 m$ height to make sure that all interactions would be recorded by the cameras. Additionally, five Estimote Location Beacons\footnote{https://estimote.com} ($B_i$) were installed into the lighting rig (high-performance, long-range beacons), configured into the device default $300ms$ \emph{Advertising Interval} and $-12dBm$ \emph{Broadcasting Power}.

Before the study began, participants were asked to read the information sheet and sign the consent form. Participants were equipped with a Radius Networks RadBeacon Dot\footnote{https://radiusnetworks.com} each (coin shaped Bluetooth~4 based low energy beacons), to place in one of their pockets. All coin-shaped beacons were pre-configured to $10ms$ advertising interval (highest) and $-18dBm$ broadcasting power (lowest), based on the results reported in Section~\ref{sec:proximity_estimation}. Half of the participants were instructed to place the beacons in the left pocket and the other half in the right pocket. The phone was always placed in the other pocket as the beacon to avoid signal interference between the two devices.

During the setup process, participants were guided through the mobile app configuration. This process included a facial photograph used to enable the later ground truth video annotations and completion of the demographic collection form for the gender, weight and height of the participant. Finally, participants were asked to synchronously perform a wave-movement in front of the cameras. The recorded sensor data of each participant was later synced with the $25fps$ video feed, achieving a sync accuracy of \(\pm\)40ms.

Participants were then instructed to socially network for a total of 45 minutes. Snacks and beverages were served before and after the end of the experiment. The discussion topic was intentionally left open, trying to simulate a realistic networking scenario. After the session, participants returned the beacons, submitted the collected data and were reimbursed with \pounds20 for their time.

In total, $99$ one-to-one interactions were observed with a mean duration of $254.9 sec$ ($\pm161.7$) and $22$ group interactions (\ie, interactions that include more than two participants) with a mean duration of $117.2 sec$ ($\pm139.4$). A separate interaction begins when the members of a group change. If the group configuration consisted less than $5 sec$, then the interaction is not counted.

All data collection and analysis was made with informed consent and approved by the ethics committee of our institution.

\subsection{Sensor Data Set}
\label{sec:case_study:Sensors}

The dataset collected for each participant contains the following sensor data:

\begin{itemize}

\item \textbf{iBeacon Proximity:} The RSSI from the mobile device with all beacons in range. This includes 24 coin-shaped beacons carried by the participants and also the high performance ceiling beacons.

\item \textbf{Linear Acceleration:} The device measured acceleration changes in three-dimensional space. This excludes the 1g acceleration produced by gravity.

\item \textbf{Gravity:} The orientation of the device relative to the ground, by measuring the 1g acceleration produced by gravity.

\item \textbf{Rotation Rate:} The device's rate of rotation around each of the three spatial axes.

\end{itemize}

The sampling rate was set to the maximum supported (100Hz) for all motion and orientation sensors. iBeacon Proximity sensor has a fixed (non-customizable) sample rate of 1Hz.

%% file: 05_Detecting_Interactions.tex
\section{Detecting Social Interactions}
\label{sec:data_analysis}

\subsection{Ground Truth}
\label{sec:data_analysis:Ground_Truth}

Video recorded from two different angles was annotated by two independent annotators using ELAN multimedia annotator software~\cite{elan}. As the aim of the study is to detect stationary interaction only, the annotators logged the beginning and end of each stationary interaction for each participant separately, using a unique ID per interaction. The annotations were cross-validated afterwards and finally verified by a third person. The instructions that the annotators followed were based on Kendon's F-formation system~\cite{Kendon:1990}:

\begin{displayquote}
\emph{An interaction begins at the moment two or more stationary people cooperate together to maintain a space between them to which they all have direct and exclusive access.}
\end{displayquote}

\subsection{Target Variable}
\label{sec:data_analysis:Target_Variable}

The dataset has a total of 645,895 labels for each combination of the 22 valid participants interacting for a total of 45 minutes in the case study. The target variable is binary, with the following two classes: {1} when a pair of participants is interacting together, and {0} when they are not. That resulted into 38,332 labels in class {1} (6.31\%), and 607,563 labels in class {0} (93.69\%).

The dataset is naturally imbalanced since it includes one label for all combinations of the participants interacting with each-other per second. The level of this imbalance obviously depends on the number of people interacting, but also on the type of interaction (\eg, one-to-one, groups of three etc.). For instance, a small event of six people will include $C(6, 2) = 15$ pairs, and thus 15 labels per second. If only three participants (A, B and C) interact together in a group of three, the dataset will include three labels of {1}, one for each combination of them (AB, AC and BC), while the remaining 12 will belong to the class {0}. In this small example, it is feasible to observe 15 labels of {1}, however as the number of participants increases this becomes impossible. Thus, lowering the feasible proportion of observed interactions from 100\% (15/15) to $\sim33\%$ ((2 x group of 9 + group of 4)/236).

\subsection{Sensor Data Pre-processing}
\label{sec:data_analysis:PreProcessing}

The data and video feed were synchronized based on the synchronous wave-movement in front of the cameras as mentioned in Section~\ref{sec:case_study:Procedure}. Each device was recording sensor data using the internal CPU time base register as timestamp, so pre-alignment between different types of sensor data (\eg, accelerometer with iBeacon Proximity) was not required. For all iBeacon Proximity data, all data reporting \emph{Unknown} values (where RSSI is $-1$) were excluded. This usually occurs at the beginning of iBeacon ranging process due to insufficient measurements to determine the state of the other device~\cite{apple_ibeacon}, or for a few seconds after the device gets out of the beacon's broadcasting range. All measurements from each user's beacon (\ie, from a participant's phone to their beacon) were also excluded. 

Since mobile devices are not real-time systems, setting a sample rate is only a suggestion to the operating system, the actual rate varies second to second. Thus, the signal for the \emph{Device Motion} sensor was re-sampled and interpolated to 100Hz. Finally, the magnitude (resultant vector) was computed from the three axes of all motion data (\ie, linear acceleration, gravity and rotation rate) available in the dataset, a process required since each user had their device in a different physical alignment and individual axis reading would not have provided useful information.

The iBeacon sensor was the only sensor that reported missing values. Since most machine learning algorithms do not accept features with unknown values, a data imputation process was required. Thus, missing values (corresponding to 5.86\% of the collected beacon data) for ceiling beacon data were imputed using linear interpolation~\cite{meijering2002chronology} on the estimated distance from the device to the ceiling beacons. Since users are changing their state less frequent, it should be possible to estimate any possible missing value reliably using this approach. For the interpersonal distances inferred from the coin-shaped beacon, missing values (corresponding to 71,88\% of the collected beacon data) were replaced with the maximum available distance, as in these cases, due to the low \emph{Broadcasting Power} that was used, the reason for missing data was that the device was out of range from the broadcaster, and thus, an interaction was not feasible.

\subsection{Proximity Estimation}
\label{sec:data_analysis:Proximity_Estimation}

The Path Loss Model (PLM) was applied in order to estimate the proximity ($d$) between each device and all beacons in range using the RSSI ($P(d)$), as shown in the following formula:

\begin{equation} \label{eq:PLM_2}
d = 10 ^ {\frac{P(d_0) - P(d) - X}{10 \times n}},
\end{equation}

where $P(d_0)$ is the Measured Power (in dBm) at 1-meter distance, $n$ the path loss exponent, $d$ the distance in which the the RSSI is estimated and $X$ a component that describes the path loss by possible obstacles between the transmitter and the receiver. The value $n = 1.5$ was set as a default constant for indoor environments~\cite{Matic:2012}. The value $X = 0$ was also chosen as it was required to measure a direct contact where no obstacles (\eg, other participants) between the two devices exist. In the situation that another participant exists in between, PLM would report a longer distance due to the decreased RSSI, and consequently, the accuracy of the distance estimation will decrease. This is a desired effect as, in the case of the coin-shaped beacons, it is only wanted to cluster whether the two users are within a range that a social interaction can be achieved. According to Hall~\cite{proxemics}, personal social interactions are achievable between 0.5 and 1.5 meters distance. Moreover, since all five long-range beacons were installed in the ceiling of the room, a clear path between the phone and most of the ceiling beacons is expected.

\subsection{Normalized Proximity}
\label{sec:data_analysis:NP}

The \emph{Normalized Proximity} (NP) is suggested by this work as an easy to compute approach for detecting social interactions using proximity-based information. More specifically, the distance of two participants is used (computed using the Path Loss Model discussed in Section~\ref{sec:data_analysis:Proximity_Estimation}) with all unknown values (\ie, when the pair is out of beacon range) being replaced with the max of all distance estimations. A proximity value $x$ is normalized into the range $[0,1]$ as follows:

\begin{equation} \label{eq:baseline}
\hat{y} = \frac{x_{max} - x}{x_{max} - x_{min}},
\end{equation}

where $\hat{y}$ is an estimate as to whether the pair is interacting, and $x$ is the estimated proximity between the pair and the $x_{min}$ and $x_{max}$ are the minimum and maximum values of $x$ for all pairs in the data set. Because $\hat{y}$ is in the range $[0,1]$ it can be compared to probability estimates.

The advantage of using this baseline compared to other works in this area (\eg, \cite{Katevas:2016:HASCA}) is that it is comparable with other probabilistic performance metrics such as Precision-Recall (PR) and Receiver Operating Characteristic (ROC) plots. Probabilistic predictions have the advantage that the designer can choose a cut-off threshold that maximizes precision or recall, depending on the use case. For example, it might be desired to choose a high precision over recall so that the model only makes a positive prediction when the probability of an interaction is very high, resulting in an accurate result with the disadvantage of losing some interactions that took place. The Normalized Proximity is also based on the estimated proximity between two people and is expected to report similar results.

\subsection{Feature Engineering}
\label{sec:data_analysis:Feature_Selection}

A series of common features were computed for all $C(22, 2) = 231$ combinations of the participant pairs. Features reflecting the current moment were initially computed, in a static window of $1 sec$, following with features reflecting past information. A set of features that are commonly included in mobile sensing problems were used, such as features extracted from motion and orientation sensors. To compute these features, the magnitude (resultant vector) of the 3-axis data were used in order to account for different physical alignment of each device within the users' pockets. Thus, no alignment of each user's motion and orientation sensors was pre-required. Based on the results from the validation experiments reported in Section~\ref{sec:proximity_estimation}, additional features were explored that provide more precise information for detecting the social interactions (\ie, interpersonal space, device position and indoor positioning features). Table~\ref{tab:features} lists the extracted features used in the data analysis of this work. The rest of this section reports on all 74 produced features and the selection strategy that was followed.

\begin{table}
\tiny
\centering
\caption{List of the features used in the data analysis.}
\label{tab:features}

\begin{tabular}{ |l|l|l| }
\hline
\multicolumn{2}{ |c| }{Table of Features} \\
\hline \multirow{2}{*}{Interpersonal Space Features}
 & $f_{prox\_rssi\_mean}$ \\
 & $f_{prox\_rssi\_diff}$ \\
\hline \multirow{4}{*}{Device Position Features}
 & $f_{device\_position\_LL}$ \\
 & $f_{device\_position\_LR}$ \\
 & $f_{device\_position\_RL}$ \\
 & $f_{device\_position\_RR}$ \\
\hline \multirow{5}{*}{Indoor Positioning Features}
 & $f_{ceiling\_beacon\_1\_diff}$ \\
 & $f_{ceiling\_beacon\_2\_diff}$ \\
 & $f_{ceiling\_beacon\_3\_diff}$ \\
 & $f_{ceiling\_beacon\_4\_diff}$ \\
 & $f_{ceiling\_beacon\_5\_diff}$ \\
\hline \multirow{7}{*}{Motion and Orientation Features}
 & $f_{time\_since\_moving\_diff}$ \\
 & $f_{device\_linear\_acc\_ccf\_lag}$ \\
 & $f_{device\_linear\_acc\_ccf\_max}$ \\
 & $f_{device\_gravity\_ccf\_lag}$ \\
 & $f_{device\_gravity\_ccf\_max}$ \\
 & $f_{device\_rotation\_rate\_ccf\_lag}$ \\
 & $f_{device\_rotation\_rate\_ccf\_max}$ \\
 \hline \multirow{8}{*}{Example of Past Information Features}
 & $f_{ceiling\_beacon\_1\_diff\_min}$ \\
 & $f_{ceiling\_beacon\_1\_diff\_max}$ \\
 & $f_{ceiling\_beacon\_1\_diff\_mean}$ \\
 & $f_{ceiling\_beacon\_1\_diff\_std}$ \\
 & $f_{time\_since\_moving\_diff\_min}$ \\
 & $f_{time\_since\_moving\_diff\_max}$ \\
 & $f_{time\_since\_moving\_diff\_mean}$ \\
 & $f_{time\_since\_moving\_diff\_std}$ \\
\hline
\end{tabular}
\end{table}

\subsubsection*{Interpersonal Space Features}

iBeacon Proximity sensor data of a pair includes two measurements: Let $rssi_{ij}$ be the RSSI between the two participants as measured from the device of user $i$ and $rssi_{ji}$ be the RSSI from the same distance as measured from the device of user $j$. The mean of the two measurements was computed as an indication of how close the two participants are in space:
\begin{equation}
f_{prox\_rssi\_mean} = (rssi_{ij} + rssi_{ji}) / 2
\end{equation}

In addition, a feature that represents the absolute difference between the two measurements was computed:
\begin{equation}
f_{prox\_rssi\_diff} = |rssi_{ij} - rssi_{ji}|
\end{equation}

Note that in this case, the raw RSSI was used as the same hardware was used for broadcasting a beacon signal across all participants, and thus, a \emph{Measured Power} constant is not required. In the case of multiple devices being used, then a feature that estimates the interpersonal distance based on a calibrated \emph{Measured Power} constant would be required, using the PLM equation mentioned in Section~\ref{sec:proximity_estimation}.

\subsubsection*{Device Position Features}

As mentioned in Section~\ref{sec:proximity_estimation}, information about the device position is important as it highly influences the RSSI signal between the two devices. For that reason, four features have been developed that includes the information of the device position (left vs. right per participant) using one-hot encoding:

\begin{itemize}

\item $f_{device\_position\_LL}$: Both P1 and P2 placed the device on the left pocket.
\item $f_{device\_position\_LR}$: P1 placed the device on the left pocket, P2 on the right.
\item $f_{device\_position\_RL}$: P1 placed the device on the right pocket, P2 on the left.
\item $f_{device\_position\_RR}$: Both P1 and P2 placed the device on the right pocket.

\end{itemize}

\subsubsection*{Indoor Positioning Features}

The absolute difference of each participant from the five ceiling beacons was computed using the following formula:
\begin{equation} \label{eq:ext_beacon_feature}
f_{ceiling\_beacon\_k\_diff} = | D_{ik} - D_{jk} |,
\end{equation}

where $D_{ik}$ is the distance reported from user $i$'s, and $D_{jk}$ is the distance reported from user $j$'s mobile device to the fixed installation $k$. It is expected that users close together would result in similar distances from the ceiling beacons and the feature will be close to zero. Note that estimated distance using PLM was used in this case instead of the raw RSSI as the long-range beacons that were used were installed in the room ceiling.

\subsubsection*{Motion and Orientation Features}

By using the measurements of the linear acceleration sensor, a feature that indicates the time since the participant has moved (in seconds) was added. A threshold of $0.15g$ was empirically chosen, indicating whether a user is moving or not, and computed the absolute difference between the pair. It is expected that if two users are moving, they will stop at the same moment and engage into a conversation, and thus, the value of that feature will be close to zero. When both users had the status `in motion', the feature was set to NaN (Unknown).

For all motion sensor data (\ie, linear acceleration, gravity, rotation rate), a cross correlation function was applied on an overlapping window of 10 seconds and extracted the maximum correlation, as well as the distance (in seconds) from the max correlation, as an indication of how similar a pair is behaving on those windows. The 10 seconds constant was chosen as indicated by \cite{Matic:2012}, but further investigation in the range of 2 to 60 also verified it as the most optimal constant. An alternative to the cross correlation function was also tested based on the Dynamic Time Warping (DTW)~\cite{salvador2007toward} method. However, due to its high computational complexity as well as its low predictive power in this context, it was excluded from the final feature list.

\subsubsection*{Past Information Features}

In order to take advantage of past information available in the data set, the \emph{min}, \emph{max}, \emph{mean} and \emph{std} was computed on all time-series features (\ie, excluding the one-hot encoded device positioning features), in an overlapping window of 10 seconds. Note that due to the length of those features, only some representative examples are listed in Table~\ref{tab:features}.

\subsection{Evaluation Procedure}
\label{sec:data_analysis:Evaluation_Procedure}

For evaluating the performance of the model, a standard 10-fold cross-validation schema was used. The dataset was initially split over time, however, due to the time-series nature of our study, a significant overfitting was reported. More particularly, since participants were changing their interactive state at any given moment, the model was memorizing the features per split and inferring them back with very high performance, due to information leakage. Thus, the data was split per participant combination (\ie, 23 samples out of 231 due to the 10-fold schema) rather than over time.

In the context of this work, \emph{Precision} is: from the detected interactions, how many of them did the model detect correctly, whereas \emph{Recall} is: from all interactions taking place, how many of them did the model detect. Depending on the use case, applications can emphasize one measure over the other. The evaluation metrics that will be used in the rest of this report is \emph{Precision-Recall (PR)} curve. Although ROC curves are heavily used when reporting performance in classification problems, due to the nature of our dataset being unbalanced, PR plots as suggested for this case by \cite{saito2015precision} and \cite{davis2006relationship} were used.

\subsection{Model Choice}
\label{sec:data_analysis:Model_Choice}

As a learning model we use XGBoost~\cite{xgboost}. XGBoost is a state-of-the-art gradient boosting regression tree algorithm that has emerged as one of the most successful feature-based learning models in recent machine learning competitions. We empirically found XGBoost consistently outperformed other well-established classifiers, such as Logistic Regression~\cite{nelder1972generalized}, Support Vector Machines~\cite{Deng:2012:SVM}, or Random Forests~\cite{Breiman:2001}. We used XGBoost v0.7.2.1 as part of the Python library scikit-learn~\cite{scikit-learn} v0.19.1 and its wrapper for the XGBoost Python package.

A parameter tuning was performed on a 20\% subset of the dataset (\ie, 46 samples out of 231). This subset was only used for the model tuning task and was never used in the training/validation procedure. The aim was to discover the model's configuration that maximizes the Average Precision (AP) performance. More specifically, a grid search algorithm over all possible combinations of the most influential parameters was followed, based on the following strategy:

\begin{itemize}

\item The total number of trees was set to 50.

\item The balance of positive and negative weights was set to: $sum(negative\_cases) / sum(positive\_cases)$,

as suggested by XGBoost documentation\footnote{http://xgboost.readthedocs.io/en/latest/parameter.html}. This was required due to the imbalanced nature of the dataset.

\item The maximum depth of the tree was tested with values [4, 6, 8, 10].

\item The number of features to consider when looking for the best split was tested with values [0.2, 0.4, 0.6, 0.8, 1].

\item The sub-sample ratio of the training instance was tested with values [0.5, 0.75, 1].

\item The model's learning rate was tested with values [0.01, 0.05, 0.1].

\item All other parameters used the library default values.

\end{itemize}

The configuration with the best performance of AP 80.4\% (\ie, performance using the 20\% subset) had the parameters {\tt max\_depth=4}, {\tt colsample\_bytree=0.2}, {\tt subsample=0.5} and {\tt learning\_rate=0.05}. This configuration is used in the rest of this section for training and validating the model with the remaining 80\% of the data set.

\subsection{Detecting Group Formations}
\label{sec:data_analysis:Group_Formations}

Detecting communities is important for a variety of applications including mobile social networks, recommender systems, security applications, and crowd management. One of our objectives is to automatically detect such group formations and classify the formed communities. Our concept for detecting group formation is based on graph theory. Each moment (in seconds) is represented as an undirected weighed graph $G = (V, E, w)$, with a set of vertices $V$ and weighed edges $E(w)$. Each vertex corresponds to a participant, each weighted edge corresponds to the probability of a pair that is interacting, as detected using the XGBoost classifier, and each detected community $C$ corresponds to a group formation.

We use a modularity optimization approach~\cite{Blondel:2008} that is fast to compute even in large networks and relies on the time-based stability of the network conditions at short time intervals~\cite{Lambiotte:2008:LaplacianDynamics}, also known as \emph{resolution parameter}. Initially, every vertex $V_i$ is assigned to a community $C_j$. Each vertex is then evaluated separately to join its neighbor's community. The join that achieves the maximum positive gain in modularity is the one that is committed. If no positive modularity is achieved, the vertex remains in its initial community. This process is applied to all vertices sequentially until it converges. Next, a new network is created using the communities as vertices ($C$), one edge between the connected communities with $C(w)$ the sum of all $E(w)$ that belong to that community, and a self-loop edge for the internal vertices. The algorithm is repeated until a maximum modularity is achieved.

We applied the community detection algorithm per second using NetworkX\footnote{https://networkx.github.io} v2.1 to handle graph operations on the network and considered a group formation when a community exists within the graph. We evaluate the performance of our approach in three ways: (a) \emph{link-level}, where a link represents an interaction between a pair of participants, (b) \emph{node-level}, where a node represents a participant that belongs to the correct interactive group, and (c) \emph{group-level}, where a group is detected to include the correct participants.

%% file: 08_Results.tex
\section{Results}
\label{sec:results}

In this section we present the results from the analysis reported in Section~\ref{sec:data_analysis}. As mentioned earlier, we report the performance of our approach in three ways: (a) \emph{link-level}, (b) \emph{node-level}, and (c) \emph{group-level}.

\begin{figure}[t!]
    \centering
    \begin{subfigure}{200pt}
        \includegraphics[width=\linewidth]{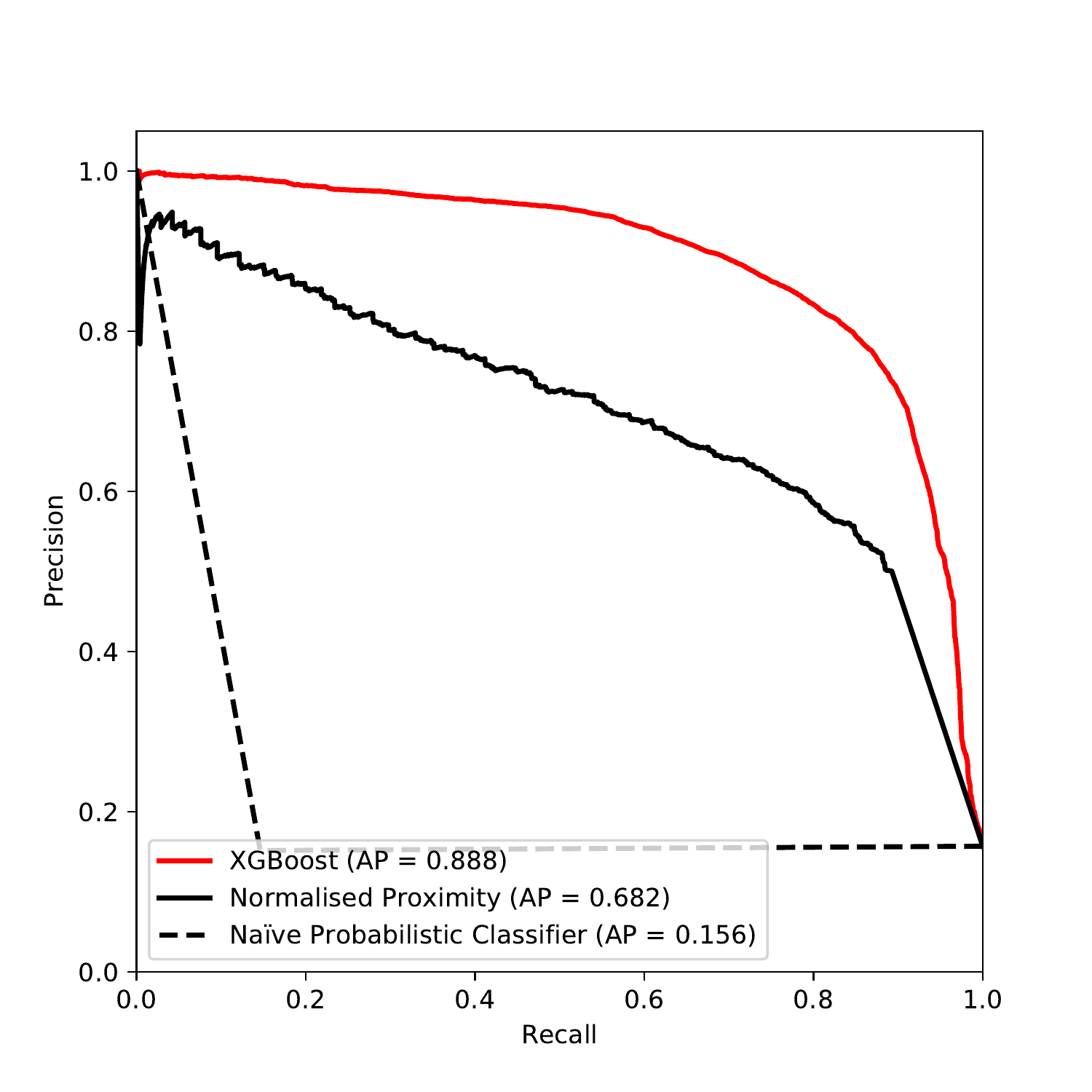}
        \caption{General performance}\label{fig:plot_performance_general}
    \end{subfigure}
    \qquad
    \begin{subfigure}{200pt}
        \includegraphics[width=\linewidth]{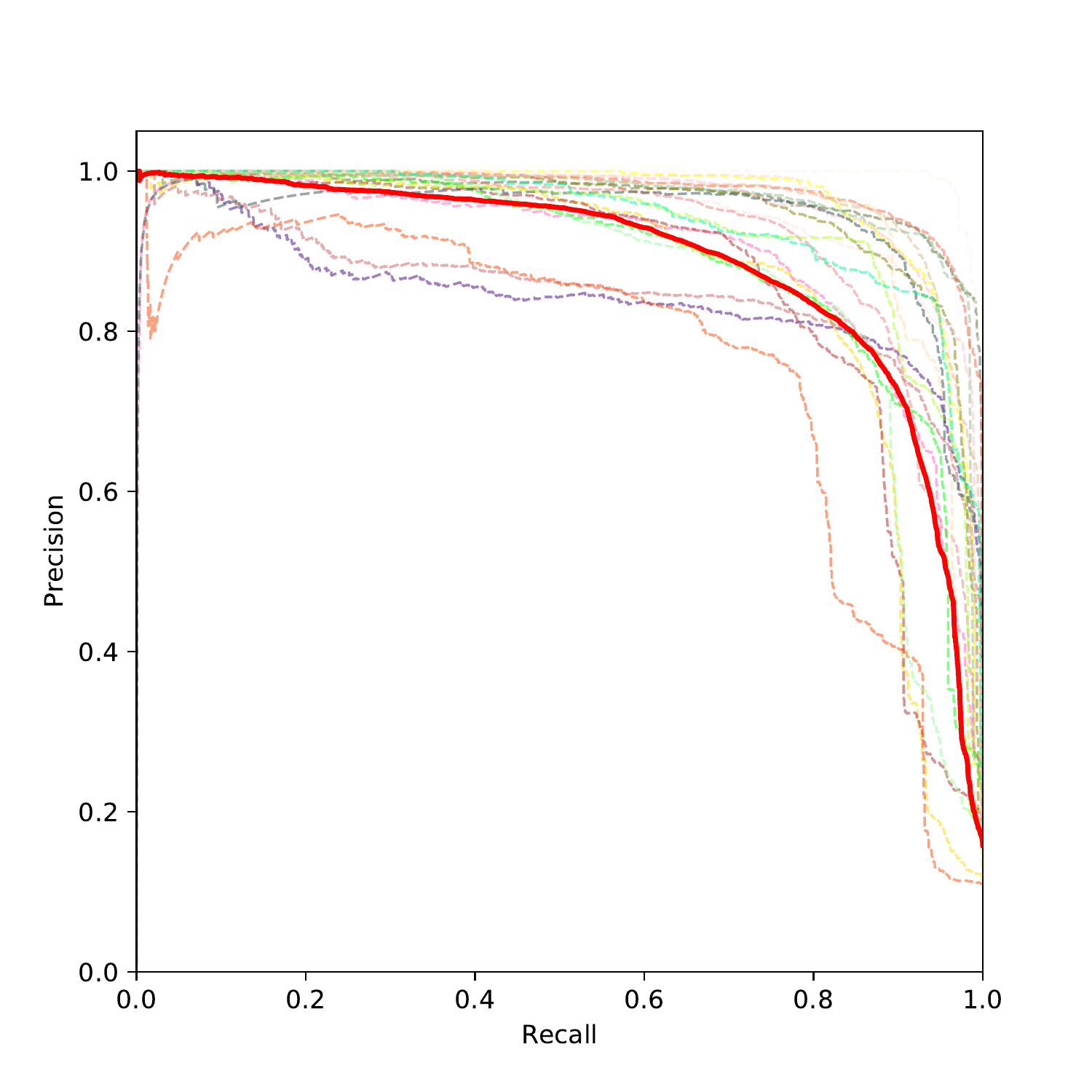}
        \caption{Performance per participant}\label{fig:plot_performance_per_participant}
    \end{subfigure}
    \caption{Performance of XGBoost classifier using a Precision-Recall (PR) Curve. The figure also includes the performance of the Na{\"i}ve Probabilistic Classifier (NPC) and the Normalized Proximity (NP) for easy comparison. The coloured lines on (b) correspond to the performance of each participant, and the thick red line corresponds to the overall average across all participants.}
    \label{fig:plot_general_performance}
\end{figure}

\subsection{Link-level Prediction}
\label{sec:Results:General_Performance}

We report the performance of the XGBoost classifier predicting the pair interactions between participants using a standard 10-fold cross-validation on the remaining 80\% samples of the dataset (\ie, excluding the 46 samples used for model tuning).

\textbf{Indoor Positioning Features --} We tested the general performance of the model, using features based on the ceiling beacons. Our aim was to detect interactions depending entirely on external infrastructure. Results reported a low performance of 18.2\%~AP, suggesting that it is not possible to achieve this with the current configuration of ceiling beacons.

\textbf{Interpersonal Distance Features --} We further tested the performance using the features related to the coin-shaped beacon. Results report a performance of 88.8\% AP (\ie, 30.2\% increase from NP and 469.2\% increase from NPC baseline). Figure~\ref{fig:plot_performance_general} shows the performance using a Precision-Recall (PR) curve plot.

Further investigation that includes both types of features (\ie, Indoor Positioning and Interpersonal Distance) reported a lower performance of 86.3\%~AP. In the rest of this work, indoor positioning features will be excluded from the analysis.

Figure~\ref{fig:plot_performance_per_participant} provides a more fine-grained analysis of the results. It depicts the user-averaged PR curve for every participant, as well as the overall performance across all participants. It is clear that the model performance varies per participant, with some of them reporting almost perfect scores, while in some others perform lower than average (red thick line).

\subsection{Sensor Importance}
\label{sec:Results:Sensor_Importance}

Accessing mobile sensor data has a significant effect on the battery life of the device, with some sensors such as the \emph{Device Motion} being one of the most power expensive sensor of all others~\cite{Katevas:2016}. In this section we explore the sensor contribution of this approach, aiming to understand which sensors produce the features that are the worst predictors and how the model's performance will be affected when they are excluded.

For measuring the sensor contribution, a leave-one-sensor-out technique was used. The model was tuned (using the same approach discussed in Section~\ref{sec:data_analysis:Model_Choice}) and then validated with all features except the ones produced by the excluded sensor. The following sensors (or external information in the case of device position) were manipulated:

\begin{itemize}

\item Interpersonal Space Features (\ie, features related to the coin-shaped beacons).
\item Device Position Features (\ie, the one-hot encoded information of the smartphone position).
\item Motion and Orientation Features (\ie, features related to the Device Motion sensor).

\end{itemize}

\begin{figure*}[t!]
    \centering
    \includegraphics[width=\linewidth,trim={0cm 1cm 0cm 0cm}]{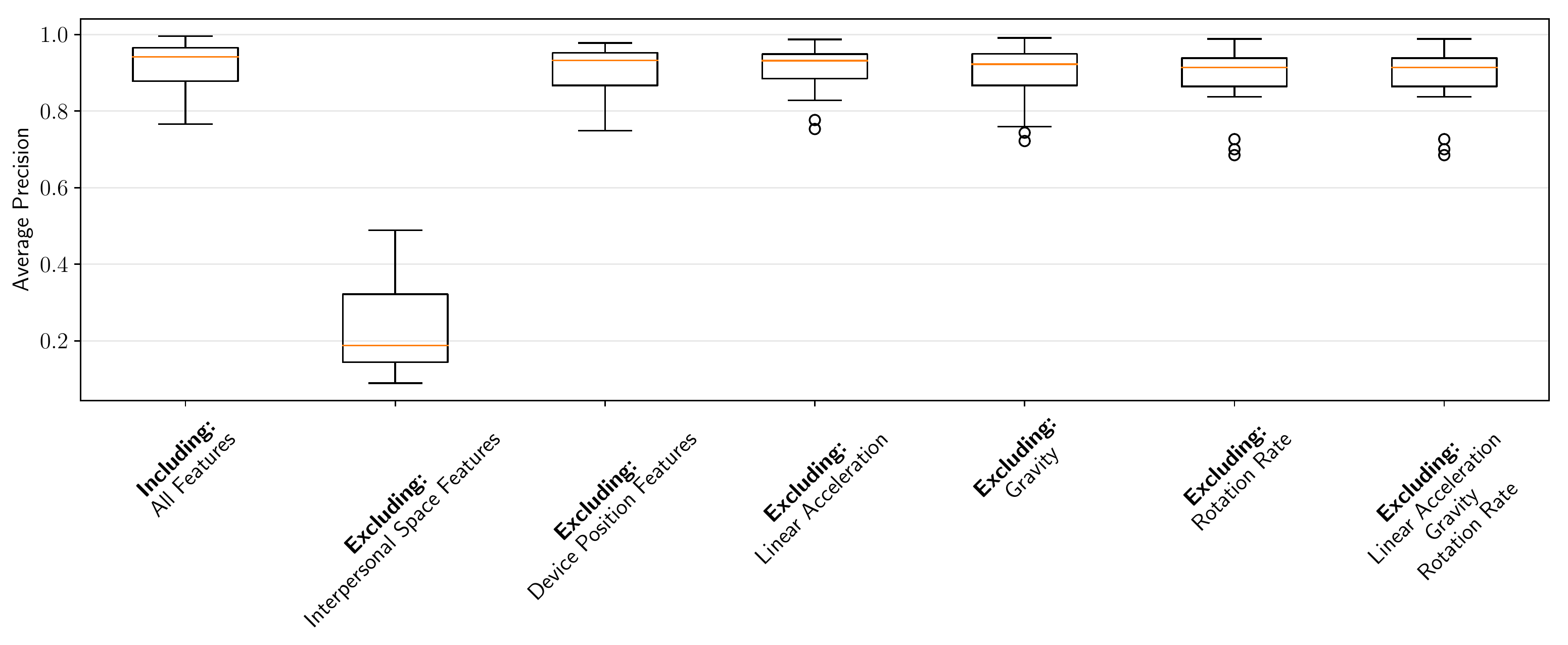}
    \caption{Sensor importance using leave-one-sensor-out as reported by the XGBoost classifier. A model that includes all features is also listed for easy comparison. The variability in the boxes corresponds to the considered participants.}
    \label{fig:sensor_importance}
\end{figure*}

In the case of \emph{Motion and Orientation Features}, the three sensor-fused data (\ie, linear acceleration, gravity and rotation rate) are explored together, but separately as well. Figure~\ref{fig:sensor_importance} shows the results from this analysis. It is evident that by excluding the interpersonal space related features the model reports random performance, similar to the NPC classifier~(\ie, AP~18.2\%). The remaining sensors have a less significant effect to the model performance, with the removal of \emph{Device Position Features} reporting AP~86.0\% and \emph{Motion and Orientation Features} reporting AP~83.0\%.

These findings suggest that a model that depends entirely on the interpersonal space features would achieve a reasonable performance, considering the fact that the contribution from the other sensors with the current engineered features is very small and might not be significant if you take into account the battery consumption of such sensors.

\subsection{Probabilistic Threshold Choice}
\label{sec:Results:Threshold_Choice}

Until now, reporting the performance of the model was made through metrics or plots that depend on probabilities (\ie, Precision-Recall as well as numerical representations of these plots such as AP). In a real-world implementation of this model, a binary prediction will be required instead of a probability. The designer of such system can choose a threshold (also called probability cut-off) of which probabilities greater or equal to this threshold would be classified as {1}, and {0} in all other cases. Choosing such threshold would lead into reporting the performance of the model in terms of \emph{precision} and \emph{recall}. Applications can emphasize one measure over the other. For example, in a use case of a mobile app that users install in order to log their interactions at a social event, the designer could emphasize on recall if the requirement is not to lose people they've interacted with, even if that results into increased false positives. If the requirement is a sticker model that captures interactions only when the probability is high, the designer could emphasize on precision.

For computing the most optimal threshold, the F1 score was used as a harmonic mean between precision and recall. Other measures such as  $F_2$ score could be used that weights recall higher than precision, or $F_{0.5}$ which puts more emphasis on precision rather than recall. F1 was maximized in the same set used for model tuning. The value $p = 0.61$ for the XGBoost model and $p = 0.48$ for the NP was found to be the most optimal that maximizes the F1 score.

Table~\ref{tab:confusion_matrix} shows the confusion matrix of the XGBoost classifier using a cut-off $p = 0.61$ and Normalized Proximity (NP) baseline with cut-off $p = 0.61$. XGBoost reports a precision of 77.8\% and recall of 86.5\%, whereas NP reports a lower precision of 61.9\% and lower recall of 74.9\%.

\begin{table}
\small
\centering
\caption{Confusion matrix for XGBoost classifier and Normalized Proximity (NP) with cut-off $p = 0.61$ and $p = 0.48$ accordingly.}
\label{tab:confusion_matrix}
\begin{tabular}{ c | c c | c || c c | c |}
    \cline{2-7}
    & \multicolumn{6}{ c | }{ Predicted Class }  \\
    \cline{2-7}
    & \multicolumn{3}{ c || }{ XGBoost } & \multicolumn{3}{ c | }{ Normalized Proximity (NP) } \\
    \cline{2-7}
              & Positive  & Negative  & Total & Positive  & Negative  & Total  \\
    \hline
    \multicolumn{1}{ | l | }{ Actual Positive } & \bf{123762 (TP)} &   5940 (FN)     & 129702 & \bf{118589 (TP)} &   11113 (FN)     & 129702  \\
    \multicolumn{1}{ | l | }{ Actual Negative } & 3269 (FP)      &  \bf{20870 (TN)} & 24139   & 6052 (FP)      &  \bf{18087 (TN)} & 24139  \\
    \hline
    \multicolumn{1}{ | c | }{ Total }    & 127031       &  26810 & 153841   & 124641       &  29200 & 153841  \\
    \hline
\end{tabular}
\end{table}

\begin{figure}[t!]
    \centering
    \includegraphics[width=300pt,trim={0cm 1cm 0cm 0cm}]{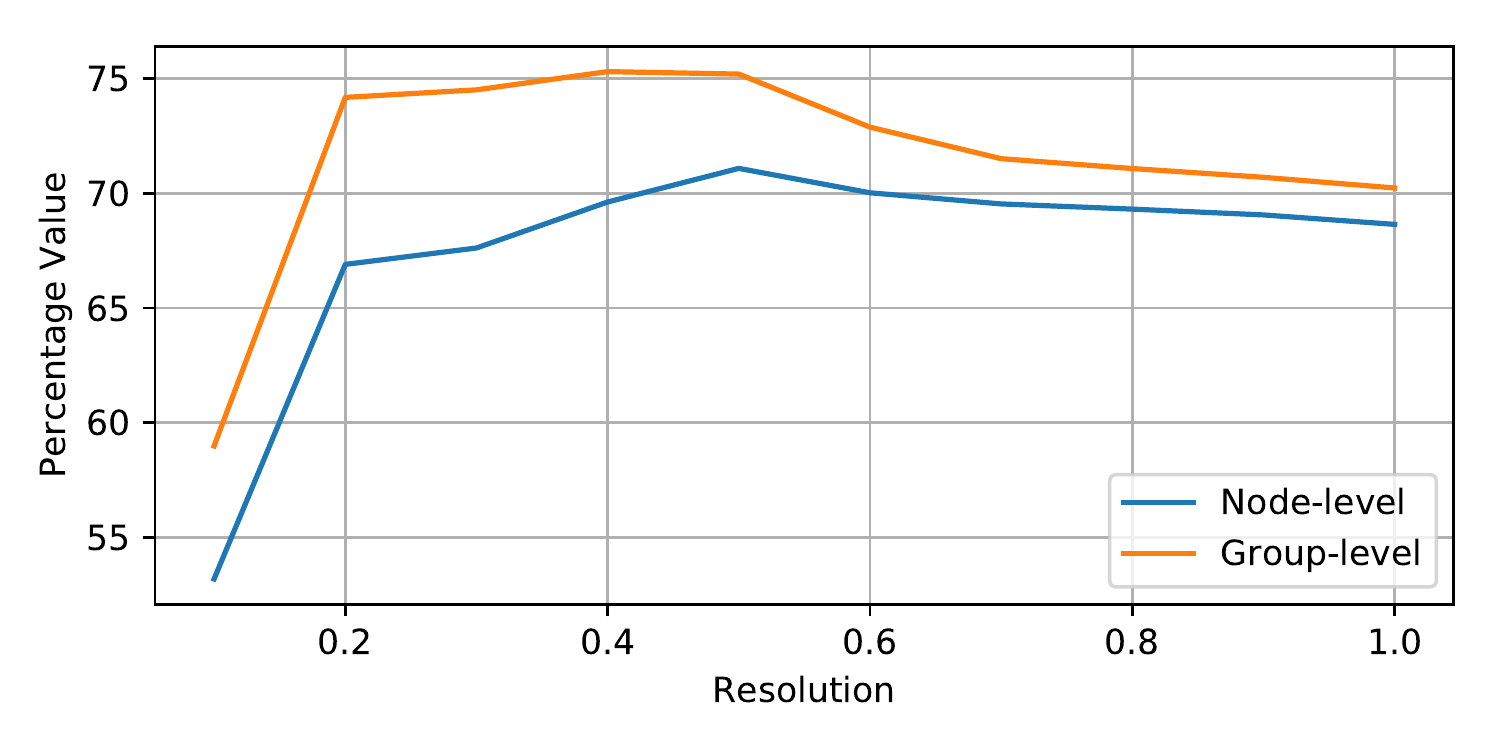}
    \caption{Performance of group formation detection at node- and group-level, using different resolution constants.}
    \label{fig:groups_performance}
\end{figure}

\subsection{Group Formation Detection}
\label{sec:Results:Group_Formation}

Figure~\ref{fig:groups_performance} shows the performance of the group detection as described in Section~\ref{sec:data_analysis:Group_Formations}. It displays the group detection accuracy on node- and group-level, using different community detection resolution constants within the range of 0.1 and 1.0.

The optimal resolution value in this case, shown in Figure~\ref{fig:groups_performance}, is $0.5$, achieving a node-level performance of 71.09\%, and group-level performance of 75.19\%. Applying the same method on the NP baseline with the optimal resolution of $0.2$ gives node-level performance of 48.65\%, and group-level performance at 50.90\%.

%% file: 09_Discussion_and_Implications.tex
\section{Discussion and Implications}
\label{sec:discussion}

Our results provide evidence that it is possible to detect interactive groups of various sizes relying on data collected from mobile devices, with a reasonable performance (77.8\% precision, 86.5\% recall and 94.0 accuracy). That means that 77.8\% of the participants that the model discovered as interacting were correctly detected, and 86.5\% of all interactions that actually took place during the event were detected by the model.
Our approach is capable of detecting group formations at node-level performance of 71.1\%, and group-level performance of 75.2\%. Moreover, our work evaluates the interactions in high granularity of 1 second windows. This is an improvement compared to other related works that binary detect if a pair has interacted during the event~\cite{Zhang:2016,Palaghias:2015}, or use longer windows of a few seconds~\cite{Montanari:2017,Matic:2012}.

\subsection{Real-World Application}

One technical challenge that arises is whether a real-time system that would continuously receive data from a larger number of participants would be possible. Such system could be implemented as a cloud-based solution that either relies on a reliable internet connection or save the data temporarily into the device and only submit when the event is completed.

As mentioned earlier, all used features are based on sensor data collected from the participants, and no external -- environment depended infrastructure (\eg, ceiling beacons) is required. The use of Bluetooth Smart sensor captured the interpersonal proximity between users, and no indoor localization techniques were used. Thus, we can assume that the model would report similar results in new environments, without requiring an additional model re-training. However, to analyze the data, features from all combinations of participants have been extracted. This is possible with a reasonable number of participants but does not scale to larger crowds. The use of ceiling beacons installed in the room, even though did not improve the performance of the final model, could be a possible solution to this limitation. By using the proximity of each ceiling beacon as a pre-filter, clustering the crowd into smaller groups and only applying the model on each cluster. Such implementation could also benefit from parallel processing, assigning clusters to analyze each region in parallel.

At the moment, such implementation is only possible using wearable beacons that simulate the beacon broadcasting of each device, due to the restrictions of current mobile operating systems mentioned earlier. Although these limitations are software restrictions from the mobile operating systems and can change in future software updates, it makes the use of an external wearable hardware essential for discovering the proximity between users, increasing the cost in a real-world implementation and application of such system.

\subsection{Battery Consumption}

As evaluated by Katevas~\etal~\cite{Katevas:2016}, collecting sensor data from mobile devices can have a noticeable impact in the device's battery. Some sensors, such as the sensor fused \emph{device motion} that reports the \emph{linear acceleration}, \emph{gravity} and \emph{rotation rate} are consuming lots of processing power. Moreover, using the internet connection to periodically submit data packets to the cloud service would have an additional impact, something that is not considered in this case.

In the current study, the battery consumption of each device was also collected through the \emph{Battery Status} sensor support of SensingKit framework~\cite{Katevas:2014}. For the total duration of the study, the battery drop was 0.08\% per hour (SD: 0.06). That included a sensor data collection from eight sensors (\ie, accelerometer, gyroscope, magnetometer, device motion, heading, iBeacon proximity and battery status) stored into the device's memory in CSV format. Note that in the current analysis, only features from the \emph{Device Motion} sensor were used from the motion and orientation sensors, sampled in the highest sampling rate of 100Hz. By decreasing the sampling rate and excluding sensors that was not used in the analysis it is expected to have a noticeable improvement in the battery consumption.

\subsection{Privacy Implications}

Even though the method depends on using anonymous IDs when broadcasting iBeacon data and no other personal information is broadcast, there is always the danger that this anonymity can be compromised by tracking the openly available ID of a user. A possible solution for protecting the user's privacy can be the use of encryption on the advertising packet, so that only authorized people or applications can make use of it. Google provides an official support of encryption in the latest Eddystone-EID frame type\footnote{https://developers.google.com/beacons/eddystone-eid}, released in April 2016. Even though not officially supported in iBeacon specification, third-party companies provide alternative solutions by rotating the beacon's attributes (\ie, Major and Minor) so that the broadcaster's ID is unpredictable\footnote{https://developer.estimote.com/ibeacon/secure-uuid/}.

Other possible privacy implications arise from the use of motion sensors (\ie, accelerometer and gyroscope). Previous works have shown that the user's identity can be exposed by capturing unique patterns from motion sensors~\cite{Shi:2011,Neverova:2016}. More recent works have successfully addressed these implications by using machine-learning-based data transformation mechanisms to hide information that can reveal the user's identity~\cite{malekzadeh2018mobile} or other sensitive activities~\cite{malekzadeh2018replacement}. In our context, a transformation model could be used in each user's phone that would pre-process the signal before submitting the data into the cloud service, preserving the user's identity.

\subsection{Limitations}

The dataset that has been analyzed, even though extended compared to other similar studies~\cite{Palaghias:2015, Katevas:2016:HASCA}, only represents a subset of what is expected in similar social gatherings, such as conferences or other networking events. Other social interactions have not been investigated such as people interacting in a coffee table, walking interactions etc.

In addition, the device position that has been tested is the trousers pocket which is a popular position according to \cite{Ichikawa:2005}. However, other positions should also be considered, such as the shoulder bags, backpacks, or even holding the device at hand.

%% file: 10_Conclusion.tex
\section{Conclusion}
\label{sec:conclusions}

In this work, we introduced a supervised machine learning approach capable of detecting stationary social interactions of a variety of sizes inside crowds. As far as we are aware, this is the first smartphone-based system that is device-type independent and capable of detecting group interactions of various sizes using mobile sensing data. Furthermore, our work does this in a relatively large (as compared to other related works) study, achieving a performance of 77.8\% precision and 86.5\% recall, when evaluating the interactions of the participants on link-level. Our approach is capable of detecting group formations at a node-level performance of 71.09\%, and group-level performance of 75.19\%. We will share our dataset that includes natural one-to-one and group interaction in varying sizes in anonymized format.

We believe that our work will be particular useful to researchers and practitioners wishing to explore crowd dynamics in social gatherings, event organizers aiming to monetize their events by providing rich analytics about their attendees, or event attendees wishing to remember their contacts without the need for exchanging business card or social media details.

Future work includes the exploration of device position detection (\ie, trousers pocket, shoulder bags and backpacks) as a way to improve the performance of the model in real-world applications. Additionally, we aim to apply a real-time version of this work in a large-scale social event and explore the ways in which the crowd is interacting in planned events.